\documentclass[aps,prl,twocolumn,superscriptaddress,preprintnumbers,amsmath,amssymb,showpacs]{revtex4}

\usepackage{graphicx,color,epsfig,rotate}
\usepackage{dcolumn}
\usepackage{bm}
\newcommand{\Tb}{TbMnO$_3$}

\newcommand{\Dy}{DyMnO$_3$}
\newcommand{\EuY}{Eu$_{0.75}$Y$_{0.25}$MnO$_3$}
\newcommand{\ir}{IR}
\newcommand{\wns}{cm$^{-1}$}

\begin{document}


\title{The origin of electromagnon excitations in multiferroic \textit{R}MnO$_3$}

\author{R. Vald\'{e}s Aguilar}
\email{rvaldes@physics.umd.edu}
 \affiliation{Department of Physics, University of Maryland, College Park, Maryland 20742}
\author{M. Mostovoy}
 \affiliation{Zernike Institute for Advanced Materials, University of Groningen, 9747 AG Groningen, The Netherlands}
\author{A. B. Sushkov}%
 \affiliation{Department of Physics, University of Maryland, College Park, Maryland 20742}%
\author{C. L. Zhang}
\affiliation{Rutgers Center for Emergent Materials and Department of Physics \& Astronomy, Rutgers University,
Piscataway, New Jersey 08854}%
\author{Y. J. Choi}
\affiliation{Rutgers Center for Emergent Materials and Department of Physics \& Astronomy, Rutgers University,
Piscataway, New Jersey 08854}%
\author{S-W. Cheong}
\affiliation{Rutgers Center for Emergent Materials and Department
of Physics \& Astronomy, Rutgers University, Piscataway, New
Jersey 08854}
\author{H. D. Drew}
\affiliation{Department of Physics, University of Maryland, College Park, Maryland 20742}

\begin{abstract}
Electromagnon excitations in multiferroic orthorhombic \textit{R}MnO$_3$ are shown to result from
the Heisenberg coupling between spins despite the fact that the static polarization arises from the much weaker
Dzyaloshinskii-Moriya (DM) exchange interaction. We present a model incorporating the structural characteristics of
this family of manganites that is confirmed by far infrared transmission data as a function of temperature and magnetic
field and inelastic neutron scattering results. A deep connection is found between the magnetoelectric dynamics of the
spiral phase and the static magnetoelectric coupling in the collinear E-phase of this family of manganites.
\end{abstract}

\maketitle

The coupling between the magnetic and ferroelectric order in a diverse set of materials termed multiferroics is currently a topic of intense study \cite{Kimura-113,Hur-nature}.
The interplay between these two orders is particularly striking in materials where ferroelectricity appears as a consequence of spontaneously breaking of inversion symmetry of the magnetic ordering. In many such magnetic ferroelectrics the spins order in an incommensurate cycloidal spiral state \cite{Cheong-Mostovoy}. The microscopic origin of ferroelectricity for this case has been discussed by a number of authors \cite{Katsura-spin-current,Sergienko-DM, Mostovoy-spiral}, leading to a consensus, generically termed the \textit{spiral mechanism}. It relies on the lowering of the energy of the antisymmetric Dzyaloshinskii-Moriya (DM) exchange in the spiral state by a polar lattice distortion, which induces an electric polarization,
$\mathbf{P}\propto \mathbf{Q} \times \mathbf{R}$, where $\mathbf{R} \propto \mathbf{S}_i \times \mathbf{S}_{i+1}$  is the spin rotation axis, and $\mathbf{Q}$ is the wave vector of the spiral. These ideas have been of central importance in the recent discovery of new multiferroic compounds.

Another apparent consequence of multiferroicity is the existence of novel coupled magnon-phonon excitations called electromagnons \cite{Pimenov-Nature,Sushkov-Y125}. A magnon that gives rise to oscillations of electric polarization can be excited by electric fields, thereby coupling much more strongly to light than the usual magnetic dipole excitation of magnons corresponding to antiferromagnetic resonance (AFMR). The resulting electric dipole spectral weight has been transferred from the phonons down to the magnon frequency. The dynamic magnetoelectric effects resulting from the coupling between spin and polarization waves were discussed theoretically at an early stage of the research on multiferroic materials \cite{Smol-Chupis}. More recently, Katsura, Balatsky and Nagaosa (KBN) noted that the magnetoelectric coupling of the \textit{spiral mechanism}, also gives rise to an electromagnon \cite{Katsura-DM}. When the spiral plane rotates around $\mathbf{Q}$, so does the induced electric polarization, which couples this magnetic excitation to electric field $e$ of a light wave normal to the spiral plane: $e \| \mathbf{R}$. The first observation of the electromagnon peak for $e \| a$ in TbMnO$_3$ with the $bc$-plane spiral spin ordering ($\mathbf{Q}\|b$) seemed to confirm this selection rule \cite{Pimenov-Nature}. However, recent measurements on other spiral multiferroics from the same family of materials, \EuY\ \cite{Rolando-EuY} and \Dy\ \cite{Kida-Dy113}, showed that this selection rule is violated  and that instead, the selection rule is tied to the lattice \cite{Andrei-review}. Therefore, the critical questions are the origin of the observed excitations, and why the KBN mechanism does not seem to apply to these multiferroics.

The violation of the KBN prediction implies that a different dynamical magnetoelectric coupling is responsible for the appearance of electromagnons in the spiral state. In this Letter we report a magneto-infrared transmission study of \Tb\ and a theoretical model based on symmetric Heisenberg exchange that clarify the origin of electromagnon excitations in cycloidal \textit{R}MnO$_3$. Additionally it is shown that this coupling is also responsible for the ferroelectricity in the collinear magnetic E-phase of rare earth manganites \cite{Sergienko-Ephase}, and based on the dynamic response of the cycloidal manganites an estimation of the polarization in the E-phase is given.

Single crystals of \Tb\ were grown as described elsewhere \cite{Rolando-Tb125}. The crystals were oriented by Laue x-ray diffraction, and surfaces polished perpendicularly to the three crystallographic axes were prepared. Two samples were used in these measurements: (1) $ab$ sample for zero field experiments of size $3\times 3$ mm$^2$ and 50 $\mu$m thick along $a$,$b$ and $c$, respectively (\textit{Pbnm} setting), and an (2) $ac$ sample for magnetic field measurements ($\mathbf{H}||b$) of size $2 \times 2 \times 1$ mm$^3$ for $e||c$ and then polished to 75 $\mu$m for $e||a$. Transmission measurements were done using a Fourier transform spectrometer connected to a magnet cryostat with optical access windows in the Faraday configuration. Sample (2) allows access to the phase where ferroelectric polarization $\mathbf{P}$ is changed from the $c$ to the $a$ axis with a field of 5 T at 7 K, and the spiral spin configuration rotates from the $bc$ to $ab$ plane. The transmission spectra were then converted to the absorption coefficient $\alpha(\nu)$ using $T(\nu) = (1-R(\nu))^2e^{-\alpha(\nu) \textit{l}}$, where $l$ is the sample thickness \cite{Heavens-Optics}.

\begin{figure}[t]
\begin{center}
\includegraphics[width=1\columnwidth]{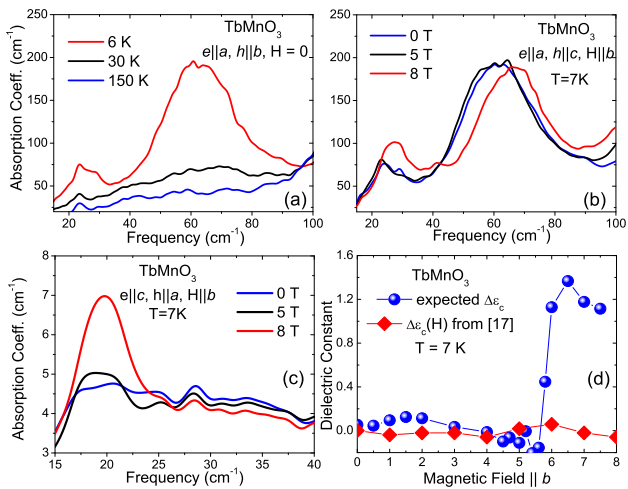}
\caption{(Color Online)(a) Zero field absorption spectra in \Tb. Main features of spectra correspond to 2 electromagnon peaks at 25 and 60 \wns\ that activate below $T_{FE}$, and the lowest infrared active phonon just below 120 \wns.
Additionally, a broad absorption below the phonon persists well above $T_N$. (b) Magnetic field dependence (\textbf{H} $\|$ \textit{b}) of the low temperature spectra. The frequencies of the electromagnons shift in the high field phase, and their intensities increase. (c) Absorption spectrum for the expected polarization of the KBN electromagnon. The new absorption line around 21 \wns\ in the high field phase is magnetic dipole active given that the expected contribution to $\varepsilon$ (shown in (d)) is larger than found in the static measurements \cite{Kimura-113-all}.}
\label{layout}
\end{center}
\end{figure}

Figure \ref{layout}(a) shows the spectra in the far infrared taken at zero external magnetic field in \Tb. All the features observed have been identified in the spectra of \EuY\ \cite{Rolando-EuY} and are described in the caption. We then performed measurements of the far infrared transmission in a magnetic field ($\mathbf{H} \| b$) to test the behavior of electromagnons on different spiral states. The spectra taken at 7 K are shown in figure \ref{layout}(b). Above approximately 5 T the system undergoes a spin flop transition from a spiral in the $bc$ plane to an $ab$ plane spiral \cite{Senff-INS-B}, yet we observe that the electromagnons remain in this configuration $e||a$ with only a slight shift of their frequencies and an increase of their spectral weight.

It is particularly striking that the absorptions activated in the cycloidal phase are only active for light polarization $e||a$ in all systems studied, regardless of the spin plane, or static polarization direction or value. This observation directly contradicts the prediction of KBN \cite{Katsura-DM}. Indeed, the inadequacy of this mechanism is also suggested by the large Born effective charge required in the KBN model to account for the experimentally observed oscillator strength.  The observed selection rule clearly indicates that account must be taken of the specific characteristics of these manganites, such as the crystal structure and anisotropic magnetic interactions, in order to find an accurate description of the electromagnon excitations.

The KBN prediction for the selection rule of electromagnons when the spiral plane rotates from $bc$ to $ab$ was tested with measurements in the $\mathbf{H} \| b$, $e\|c$, $h\|a$ configuration. The absorption spectra are shown in Fig.~\ref{layout}(c). The spectrum in the high field phase shows clearly an absorption feature around 21 \wns. By fitting the transmission spectrum with a model electric dipole active excitation we estimate the putative contribution to the static dielectric function $\varepsilon$ as $\Delta\varepsilon \approx 1.2$ as shown in figure \ref{layout}(d). This change in $\varepsilon$ is not observed in the static data \cite{Kimura-113-all}, showing that this feature is not an electric dipole excitation. Therefore this excitation is a magnetic dipole ($h\|a$) AFMR as expected from the magnetic order. For comparison we note that in the zero field spectra in the $e\|b$, $h\|c$ configuration an AFMR around 22 \wns\ is observed (not shown). This means that the $h\|c$ magnon rotates together with the spin plane thereby changing its selection rule to $h\|a$, consistent with the change of polarization direction \cite{Kimura-113} and recent neutron scattering measurements \cite{Senff-INS-B}.

\begin{figure}[t]
\begin{center}
\includegraphics[width=0.65\columnwidth]{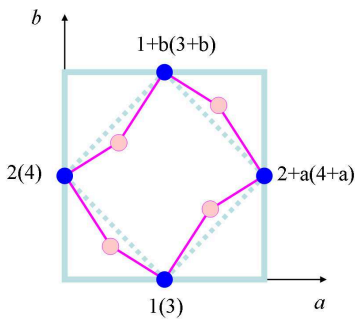}
\end{center}
\caption{(Color Online) One $ab$ layer of the \textit{Pbnm} unit cell of
\textit{R}MnO$_3$ consisting of 4 Mn ions (blue circles) and
4 oxygen ions (pink circles). The displacements of the oxygen ions
from the midpoints of the straight lines (dashed lines) connecting
neighboring Mn ions are $\delta$r$_{1+b,2+a}$ = ($\delta$x;
$\delta$y; $\delta$z), $\delta$r$_{1,2}$ = ($-\delta$x; $-\delta$y;
$-\delta$z), $\delta$r$_{1,2+a}$ = ($-\delta$x; $\delta$y;
$\delta$z), and $\delta$r$_{1+b,2}$ = ($\delta$x; $-\delta$y;
$-\delta$z). The labels of the Mn ions in the next layer (with $z = c/2$) are
given in parentheses. The displacements of the oxygen ions in the next
layer have opposite $\delta$z.} \label{distortion}
\end{figure}

The essential features of magnetic excitations in the spiral state can be understood within a model including only superexchange interactions between Mn spins,
\begin{equation}
H_{\rm ex} = \frac{1}{2} \sum_{i,j} J_{ij}\,{\bf S}_{i} \cdot {\bf S}_{j}.
\label{eq:superexchange}
\end{equation}
When magnetic anisotropies are neglected, the 4 Mn ions in the unit cell of \textit{R}MnO$_3$, located at ${\bf R}_{1}
= (\frac{1}{2},0,0)$, ${\bf R}_{2} = (0,\frac{1}{2},0)$, ${\bf R}_{3} = (\frac{1}{2},0,\frac{1}{2})$, and ${\bf R}_{4}
= (0,\frac{1}{2},\frac{1}{2})$, become magnetically equivalent (see Fig.~\ref{distortion}). The competition between the nearest-neighbor ferromagnetic exchange, $J = J_{i,i+(b \pm a)/2} < 0$, and the antiferromagnetic next-nearest-neighbor exchange along the $b$ axis, $J_{b} = J_{i,i+b} > 0$, favors the circular spin spiral with the wave vector $Q\|b$, where $\cos \frac{Q}{2} = \frac{|J|}{2J_{b}}$, provided that $J_{b} > \frac{|J|}{2}$ and $J_{a} = J_{i,i+a} <\frac{J^2}{4J_{b}}$. The antiferromagnetic exchange along the $c$ axis, $J_{c} = J_{i,i+c/2} > 0$, gives rise to a `double spiral' structure with antiparallel spins in neighboring $ab$ layers:
\begin{equation}
\left\langle {\bf S}_{\rm i}\right\rangle = \pm \left({\hat {\bf c}} \cos {\bf Q}\!\cdot\!{\bf r}_{\rm i}-{\hat {\bf
b}} \sin {\bf Q}\!\cdot\!{\bf r}_{\rm i}\right),
\label{eq:aveageS}
\end{equation}
where the upper (lower) sign corresponds the $ab$ layers with integer (half-integer) $z/c$. The $bc$ plane is favored e.g. by the single-ion anisotropy, $\frac{K}{2} \sum_{i} \left(S_{i}^{a}\right)^2$ with $K > 0$, which does not spoil the equivalence of Mn sites (effects of other magnetic anisotropies are discussed below).

\begin{figure}[t]
\begin{center}
\includegraphics[width=0.8\columnwidth]{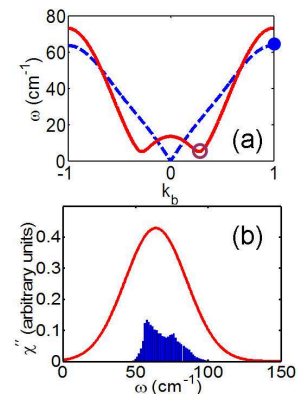}
\end{center}
\caption{(Color Online) (a) The typical magnon spectrum for a circular $bc$-spiral
along ${\bf k} = (0,k_b,0)$, where $k_b$ is measured in units of
$\frac{2\pi}{b}$. The empty and filled circles mark the position of,
respectively, the KBN electromagnon and the electromagnon excited
through the rotationally invariant coupling Eq.(\ref{eq:coupling}).
(b) Corresponding dielectric susceptibility for the (arbitrarily broadened) single-magnon peak (solid line) and
bi-magnon spectrum (histogram).} \label{fig:magnonspectrum}
\end{figure}

The magnon spectrum of the `double spiral' state has one acoustical and one optical branch plotted in Fig.~\ref{fig:magnonspectrum}(a) with, respectively, blue dashed and red solid line, as a function of the wave vector $\mathbf{k}$ in the co-rotating spin frame, in which $\langle {\bf S}\rangle \| {\hat z}$ on all sites. For acoustical magnons the order parameter ${\bf L} = {\bf S}_1 + {\bf S}_2 - {\bf S}_3 - {\bf S}_4$ oscillates in the spiral plane, while optical magnons correspond to out-of-plane oscillations of ${\bf L}$. The KBN electromagnon is the symmetric superposition of the optical magnons with the wave vectors $\pm\mathbf{Q}$, while the antisymmetric superposition is the AFMR discussed above.

Next we discuss a mechanism of magnetoelectric coupling that only involves the isotropic Heisenberg exchange between non-collinear spins  and, therefore, is insensitive to the orientation of the spiral plane.  Due to the GdFeO$_3$ distortion of \textit{R}MnO$_3$ compounds, oxygen ions mediating the superexchange between nearest-neighbor spins in the $ab$ layers are displaced from the straight lines connecting two neighboring Mn ions in $ab$ layers as shown in Fig.~ \ref{distortion}. When an applied electric field shifts all oxygen ions by an equal distance along the \textit{a} axis, the exchange constants $J_{1,2}$ and $J_{1,2+a}$ will be changed by $\Delta J$ proportional to the applied field, while the exchange constants $J_{1+b,2}$ and $J_{1+b,2+a}$ will be changed by $-\Delta J$. In other words, due to the alternating rotations of the MnO$_6$ octahedra in \textit{R}MnO$_3$, a uniform electric field in the \textit{a} direction gives rise to an alternation of the nearest-neighbor exchange along the spiral propagation vector $Q\|b$, $J\propto J_0 + \Delta J$cos$(\mathbf{k_0} \cdot {\bf r})$, where $\mathbf{k_0}$ = (0, $\frac{2\pi}{b}$,0). The corresponding coupling of spins to the electric field $e\|a$ compatible with the \textit{Pbnm} symmetry has the form:
\begin{eqnarray}
H_{\rm me} = &-& g E_{a} \sum_{j} \left[ \left(\mathbf{S}_{1,j}
-\mathbf{S}_{1,j+b}\right) \cdot \left(\mathbf{S}_{2,j} +
\mathbf{S}_{2,j+a}\right) \right. \nonumber \\ &+& \left.
\left(\mathbf{S}_{3,j} - \mathbf{S}_{3,j+b}\right) \cdot
\left(\mathbf{S}_{4,j} + \mathbf{S}_{4,j+a}\right)\right],
\label{eq:coupling}
\end{eqnarray}
where the indices 1 -- 4 label different Mn ions in a unit cell while $j$ labels cells.

This interaction couples electric field through the alternation of the Heisenberg exchange along the \textit{b} axis to the acoustical magnon with $\mathbf{k}_{0}$ at the boundary of the magnetic Brillouin zone of the model Eq.(\ref{eq:superexchange}). This excitation corresponds to relative rotations of the spins ${\bf S}_{1}$ and ${\bf S}_{2}$ in the spiral plane, which occur in anti-phase with the rotations of the spins ${\bf S}_{3}$ and ${\bf S}_{4}$ and result in alternation of the angle between neighboring spins along the spiral propagation vector.

This mechanism of dynamic magnetoelectric coupling works only for non-collinear spins because the first-order variation of the exchange energy $\propto \left(\mathbf{S}_{i} \cdot \mathbf{S}_{j}\right)$ with respect to spin oscillations $\delta \mathbf{S}_{i}(t) = (\mathbf{S}_{i}(t) - \langle \mathbf{S}_{i}\rangle) \perp \langle \mathbf{S}_{i} \rangle$ is zero for collinear spins. Although interactions of spins with $e\|b$ and $e\|c$, similar to Eq.(\ref{eq:coupling}), are allowed by symmetry, they give rise to alternation of the nearest-neighbor exchange constants in the directions orthogonal to the spiral wave vector $\mathbf{Q}$ and do not lead to single-magnon excitations by electric field, as neighboring spins along the \textit{a} and \textit{c} axes are collinear. This explains why the polarization of electromagnon in \textit{R}MnO$_3$ is parallel to the \textit{a} axis independently of the orientation of the spiral plane.

We note that the coupling Eq.(\ref{eq:coupling}) induces a static electric polarization $P_{E} \| a$ in \textit{R}MnO$_3$ with the collinear antiferromagnetic ordering of the E-type \cite{Sergienko-Ephase}, for which the scalar product of neighboring spins alternates along the \textit{b} axis. This fact allows us to estimate the value of the polarization in the E-phase using the optical absorption data for the spiral phase (the direct measurement of $P_{E}$  is difficult due to absence of single crystal samples). The total spectral weight of the 60 \wns\ peak, $S \approx 7000$ cm$^{-2}$,  corresponds to $P_{E} \sim 1 \mu\!\!$ C/cm$^{2}$, in perfect agreement with \textit{ab initio} calculations \cite{Picozzi-dual}. For comparison, the spectral weight of the KBN electromagnon, calculated from the polarization $P \sim 8\times 10^{-2} \mu\!\!$ C/cm$^{2}$ in the spiral phase of TbMnO$_3$, is only  $\approx 10$ cm$^{-2}$. The photoexcitation of magnons through the magnetoelectric coupling Eq.(\ref{eq:coupling}) of exchange origin, which is about 2 orders of magnitude stronger than the relativistic coupling in spiral manganites, is much more effective than the KBN mechanism.

The coupling Eq.(\ref{eq:coupling}), as well as another invariant, $E_{a}\left[ \left(\mathbf{S}_{1} -\mathbf{S}_{1+c}\right) \cdot \mathbf{S}_{3}\right.$ -- $\left.\left(\mathbf{S}_{2} - \mathbf{S}_{2+c}\right) \cdot \mathbf{S}_{4}\right]$, gives rise to the photo-excitation of a pair of magnons with the total wave vector $\mathbf{k}_{0}$(the so-called `charged magnons' \cite{Damascelli-charged}). The calculated shape of the bi-magnon continuum for the coupling Eq.(\ref{eq:coupling}) is shown in Fig.~\ref{fig:magnonspectrum}(b). Its maximum is located around the electromagnon frequency and its total spectral weight for all reasonable parameters of the model is less than $10\%$ of the spectral weight of the single-magnon peak. This shows that the peak at 60 \wns\ results from the photoexcitation of a single zone boundary magnon, which can be identified in the magnon spectra obtained by inelastic neutron scattering \cite{Senff-review}. Even though charged magnons can also be excited by $e \|c$ and $e\|b$ through couplings similar to Eq.(\ref{eq:coupling}), strong absorptions in these polarizations are not observed in the experiments \cite{Rolando-EuY}.

Concerning the nature of the weaker 25 \wns\ peak, we note that for an elliptical spiral structure the magnon wave vector ${\bf k}$ is not conserved and a number of different mechanisms couple high- and low-frequency magnons. Thus, the spiral ellipticity resulting from the magnetic anisotropy in the spiral plane gives rise to satellite single-magnon peaks and bi-magnon continua at the total wave vector ${\bf k}_{0}\pm 2{\bf Q}$, ${\bf k}_{0}\pm 4{\bf Q} \ldots$, which could correspond to the low frequency peak. Furthermore, there are anisotropic spin-spin interactions, which are insensitive to the flop transition (e.g. involving products $S_{i}^{a} S_{j}^{c}$) and couple the electromagnon with ${\bf k} = {\bf k}_0$ to magnons with ${\bf k} = \pm {\bf Q}$ giving rise to low frequency absorption. Which of these mechanisms gives the dominant contribution to the low-frequency absorption is not clear, but all of them are different from the coupling inducing electric polarization in the spiral state.

In conclusion, we have presented experimental data and a theoretical model that strongly support the Heisenberg exchange interaction as the origin of the magnetoelectric dynamics in the family of multiferroics \textit{R}MnO$_3$. Additionally, the fact that these excitations are observed both in the neutron measurements and the \ir\ experiments reflects their origin as single-particle excitations of a hybrid electromagnon nature. The surprising outcome of this study is that optical data can be used to explore properties of competing ferroelectric states: from the measured spectral weight of the electromagnon peak in the spectrum of spiral manganites we obtained the value of the spontaneous polarization in manganites with the E-type collinear ordering. Our results imply that the static and dynamic magnetoelectric coupling are different in general and therefore that electromagnons can be observed, in principle, in non-multiferroic materials with non-collinear spin orders.  This observation opens a new avenue of investigation of the dynamic properties of frustrated magnets.

\begin{acknowledgments}
This work was supported by the NSF MRSEC program under grant DMR-0520471. MM was supported by the Stichting voor
Fundamenteel Onderzoek der Materie (FOM).
\end{acknowledgments}

\noindent{\it{Note added}}: During the final preparation of this manuscript we became aware of the work of \citet{Takahashi-Tb113} where they report very similar results in \Tb. However, they interpret the excitations as coming from an electric dipole continuous band of two-magnon absorption, which our work in fact rules out.

\bibliography{origin113mac}
\end{document}